\documentstyle[12pt,epsf]{article}

\begin{document}

\newcommand{\sheptitle}
{Relic Dark Energy from Trans-Planckian Regime}

\newcommand{\shepauthor}
{Laura Mersini, Mar Bastero-Gil and  Panagiota Kanti}

\newcommand{\shepaddress}
{Scuola Normale Superiore and INFN, Piazza dei Cavalieri 7,\\
I-56126 Pisa, Italy}

\newcommand{\shepabstract}
{}

\begin{titlepage}
\begin{flushright}
hep-ph/0101210\\
SNS-PH/01-01\\
\today
\end{flushright}
\vspace{.1in}
\begin{center}
{\large{\bf \sheptitle}}
\bigskip \medskip \\ \shepauthor \\ \mbox{} \\ {\it \shepaddress} \\
\vspace{.5in}

\bigskip \end{center} \setcounter{page}{0}
\shepabstract
\begin{abstract}

As yet, there is no underlying fundamental theory for the
transplanckian regime. There is a need to address the issue of how the
observables in our present Universe are affected by processes that may
have occurred at superplanckian energies (referred to as the
{\it transplanckian regime}). Specifically, we focus on the impact the
transplanckian regime has on two observables, namely: dark energy and
the CMBR spectrum.  We model the transplanckian regime by introducing
a 1-parameter family of smooth non-linear 
dispersion relations which modify the frequencies at very short
distances.
A particular feature of the family of dispersion functions chosen  is
the production of 
ultralow frequencies at very high momenta $k$ (for $k>M_P$). We name
the range of the ultralow energy modes (of very short distances) that
have frequencies equal or less than the current Hubble rate $H_0$ as
the {\it tail} modes. These modes are still frozen today due to the
expansion of the Universe. We calculate their energy today and show
that the $tail$ provides a strong candidate for the {\it dark energy}
of the Universe. During inflation, their energy is about 122-123 
orders of magnitude smaller than the total energy, for any random
value of the free parameter in the family of dispersion relations. 
For this family of dispersions, we present the exact solutions and
show that: the CMBR spectrum is that 
of a (nearly) black body, and that the adiabatic vacuum is the only
choice for the initial conditions.  
\end{abstract}

\vspace*{3cm}

\begin{flushleft}
\hspace*{0.9cm} \begin{tabular}{l} \\ \hline {\small Emails:
mersini@cibs.sns.it, bastero@cibs.sns.it, kanti@cibs.sns.it }
\end{tabular}
\end{flushleft}

\end{titlepage}

\section{Introduction} 
There is still no fundamental physical theory of the very early
universe which addresses issues that arise from the regime of
transplanckian physics.  One of these issues relates to the origin of
the cosmological perturbation spectrum. In an expanding Universe, the
physical momentum gets blue-shifted back in time, therefore the
observed low values of the momentum today that contribute to the CMBR
spectrum may have originated from values larger than the Planck mass
$M_P$ in the Early Universe. This is similar to the problems that
arise in trying to explain the origin of Hawking radiation in Black
Hole physics.  
In a series of papers \cite{jacobson, unruh, brout, hambli,
jacobson2}, it was demonstrated that the Hawking radiation remains
unaffected by modifications of the ultra high energy regime, expressed
through the modification of the usual linear dispersion relation at
energies larger than a certain ultraviolet scale $k_C$. Following a
similar procedure, in the case of an expanding
Friedmann-Lemaitre-Robertson-Walker (FLRW) spacetime, Martin and
Brandenberger  in Ref. \cite{brand} (see also
\cite{blhu,niemeyer,kowalski,kempf,easther})  
showed that standard predictions of inflation are indeed sensitive to 
trans-planckian physics: different dispersion relations lead to
different results for the CMBR spectrum.

It is the lack of a fundamental theory, valid at all energies,  that
makes the model building of the transplanckian regime very
interesting. The main issue is how much are the known
observables affected by the unknown theory. 
The apparently {\it ad hoc} modification of the dispersion relation
at high energies is contrained by the criterion that its low energy
predictions do no conflict the observables. 
Specifically, in this
paper we address two questions: a) can the transplanckian regime
contribute to the dark energy of the universe, and b) how sensitive is
the CMBR spectrum to energies higher than the Planck scale $M_P$,
where our current physics theory is known to break down. 

We choose a family of dispersion relations for the frequency of the
wavefunctions 
that modifies the behaviour of the field at the ultrahigh energies of
the transplanckian regime. The dispersion relation has the following
features: it is smooth, nearly linear for energies less than the
Planck scale, reaches a maximum, and attenuates to zero at ultrahigh
momenta thereby producing ultralow frequencies at very short
distances. We name the $tail$ that part of the dispersion graph of
very short distances that contains the range of ultralow frequencies
less or equal to the current Hubble constant $H_0$. It follows that 
the $tail$ modes are still currently frozen. We calculate the
energy of the $tail$ modes in order to address the former question (a)
and show that although the $tail$ does not contribute significantly to
the CMBR spectrum, it has a dominant contribution to the dark energy
of the universe \cite{dark}. The energy density of the $tail$ modes is
of the same order today as the matter energy density.

The second question (b) is motivated by the problem that in most
inflationary models the present large scale structure of the Universe
is extrapolated from a regime of ultra-high energies (known as the
transplanckian regime) originating from before the last 60 e-foldings
of the exponential expansion. In Refs. \cite{brand,niemeyer} the authors have
demonstrated that the  problem of calculating the spectrum of
perturbations with a time-dependent dispersive frequency 
can be reduced to the familiar topic of particle creation on a
time-dependent background \cite{partcreation}. We will use their
observation in what follows. They also conjecture that the observed
power spectrum can always be recovered only by using a smooth
dispersion relation, which 
ensures an adiabatic time-evolution of the modes. By taking the
frequency dispersion relations to be the general class of Epstein
functions \cite{epstein}, we check and lend strong support to their
conjecture. We present the exact solutions to the mode equation for
the scalar field\footnote{These functions are known for having exact
solutions to second order differential equations in terms of hypergeometric
functions.} with a ``time-dependent mass'',  and the resulting CMBR
spectrum below. We show that the major contribution to the
CMBR spectrum comes from the long wavelength modes when they re-enter
the horizon. The spectrum is nearly insensitive to the very short
wavelength modes inside the Hubble horizon.
 
The paper is organized as follows: in Section 2, we present the set-up
and formalism of our analysis. 
The family of dispersion functions, exact solutions to the mode
equations of motion and the resulting CMBR spectrum (from the
Bogoliubov method) are reported in Section 3. 
In Section 4, we calculate
the contribution of the {\it tail} modes to the dark energy of the
universe today. 
In this work, we have neglected the backreaction effects of particle
production.  This assumption is fully justified from the calculation of
the energy for the transplanckian modes, in Section 4. Due to the 
{\it dispersed}
ultralow frequency of these modes, the energy contained in that
transplanckian regime is very small ($10^{-122} \rho_{total}$), thus the
backreaction effect is reasonably negligible \cite{tanaka, brand}. 
We present our conclusions in Section 5.

\section{The Set-Up and Formalism} 

Let us start with the generalized Friedmann-Lemaitre-Robertson-Walker
(FLRW) line-element which, in the presence of scalar and tensor
perturbations, takes the form \cite{FLRW1, FLRW2}
\begin{eqnarray}
ds^2&=& a^2(\eta) \left\{ -d\eta^2 + \left[ \delta_{ij} + h(\eta,{\bf
n})Q \delta_{ij} \right. \right.  \nonumber \\ & &\left. \left. + h_l
(\eta,{\bf n}) \frac{Q_{ij}}{n^2} + h_{gw}(\eta,{\bf n}) Q_{ij}
\right] d x^i d x^j \right\} \,,
\label{frw}
\end{eqnarray}
where $\eta$ is the conformal time and $a(\eta)$ the scale factor. The
dimensionless quantity ${\bf n}$ is the comoving wavevector, related
to the physical vector ${\bf k}$ by ${\bf k}= {\bf n}/a(\eta)$ as
usual. The functions $h$ and $h_l$ represent the scalar sector of
perturbations while $h_{gw}$ represents the gravitational waves.
$Q(x^i)$ and $Q_{ij}(x^i)$ are the eigenfunction and eigentensor,
respectively, of the Laplace operator on the flat spacelike
hypersurfaces. For simplicity, we will take a scale factor $a(\eta)$
given by a power law\footnote{It has been argued in \cite{brand} that
the analysis extends to other laws for the scale factor.},
$a(\eta)=|\eta_c/\eta|^{\beta}$, where $\beta \geq 1$ and $|\eta_c|=
\beta/H(\eta_c)$. The initial
power spectrum of the perturbations can be computed once we solve the
time-dependent equations in the scalar and tensor sector. The mode
equations for both sectors reduce \cite{mode1, mode2,equations} to a
Klein-Gordon equation of the form
\begin{equation}  
\mu_n^{\prime \prime} + \left[ n^2 - \frac{a^{\prime \prime}}{a} \right]
\mu_n=0 \,,
\label{kg}
\end{equation}
where the prime denotes derivative with respect to conformal time.
Therefore, studying perturbations in a FLRW background is equivalent
to solving the mode equations for a scalar field $\mu$ related
(through Bardeen variables \cite{equations}) to the
perturbation field  in the expanding
background. The above equation represents a linear dispersion relation
for the frequency $\omega$,
\begin{equation}
\omega^2= k^2 = \frac{n^2}{a^2} \,.
\label{ldis}
\end{equation}
The dispersion relation of Eq. (\ref{ldis}) holds for values of
momentum smaller than the Planck scale. There is no reason to believe
that it remains linear at ultra-high energies larger than $M_P$.  Yet,
nonlinear dispersion relations are quite likely to occur from the
Lagrangian of some effective theory obtained by the yet unknown
fundamental theory. Nonlinear dispersion relations, similar to the
ones we consider in this work, are known to arise in effective
theories of: nonlocal condensed matter or particle physics models
arising from non-canonical kinetic terms \cite{nckinetic1,nckinetic2}; 
from the  dissipative behavior of a quantum system immersed into an
environment after coarse-graining \cite{cg}; or from effective
theories with phase transitions, time-dependent mass squared terms or
effective potentials \cite{effpot,kolb,linde}.  Perhaps,
trans-planckian models motivated by superstring theory
\cite{superstring1,superstring2} or a two-stage 
inflationary model \cite{twoinflation} are plausible.  In the latter
case, one could easily envision 
for example a scenario with the first stage of inflation occurring at
energy scales above the Planck mass~\footnote{There is no reason why
inflation must only occur below Planck energies. In principle,
inflation at ultra high energies is equally possible.} followed by a
nonthermal phase transition \cite{nonthermal}.  The 
preheating \cite{nonthermal, preheating} from the nonthermal phase
transition then leads to the second stage of inflation below Planck energies.
In the former case, the motivation comes from the
common belief that the superstring theory is the one that describes or
at least is valid at energies of the transplanckian period. Taking
this idea one step further, we incorporate the concept of superstring
duality (which applies at transplanckian regimes) in our analysis by
choosing a particular family of dispersion relations that exhibits
{\it dual} behavior\footnote{For example, when compactifying superstring
theory in a torus topology, of large radius $R$ and winding radius $r$, the
frequency mode spectrum is dual in the sense that $R$ and $r$ are related as
$r=1/R$. This means that each normal mode with a frequency $n/R$,
where $n$ is an integer, 
has its dual winding mode with decreasing energy that goes like
$1/r=R$ \cite{superstring2}.}, 
 i.e. appearance of ultra-low mode frequencies
both at low and high momenta\footnote{We would like to thank
A.  Riotto for pointing this out to us.} $k$.

Despite the above comments and possible approaches, we should
stress that any modeling of Planck scale physics by analogy with the
already familiar systems is pure speculation. We lack the fundamental
theory that may naturally motivate or reproduce such dispersion
behaviour. Nevertheless, it would be instructive to derive these
dispersion relations from particle physics and string theory, as a
step towards understanding the physical nature of the model.

In what follows, we replace the linear relation $\omega^2(k) = k^2
=n^2/a(\eta)^{2}$ with a nonlinear dispersion relation
$\omega(k)=F(k)$. The family of dispersion functions $F(k)$ for our
model is introduced in section 3. These functions 
have the following features: they are linear for low momenta up to 
the Plank scale $k_C=M_P$, taken to be the cuttoff scale, but beyond
the cutoff they smoothly turn down and asymptotically approach zero
whereby producing ultra-low frequencies at very short distances. 
Therefore, in Eq.  (\ref{kg}), $n^2$ should be replaced by:
\begin{equation}
n_{eff}^2 = a(\eta)^2 F(k)^2 = a(\eta)^2 F[n/a(\eta)]^2 \,.
\label{neff}
\end{equation}
We will also consider the general case of non-minimal or conformal
coupling to gravity by keeping some arbitrary, unspecified, coupling
constant $\xi$. Then, the equation for the scalar and tensor
perturbations, that we need to solve, takes the form
\begin{equation}
\mu_n^{\prime \prime} + \left[n^2_{eff} - (1- 6\xi)\frac{a^{\prime
\prime}}{a}\right] \mu_n = 0 \,.
\label{modeneff}
\end{equation}
For future reference, we define the generalised comoving frequency as
\begin{equation}
\Omega_n^2= n^2_{eff} - (1- 6\xi)\frac{a^{\prime \prime}}{a}\, \label{omegan}.
\end{equation}
The dynamics of the scale factor is determined by the evolution of the
background inflaton field $\phi$, with potential $V(\phi)$, and the
Friedmann equation. In conformal time, these equations are:
\begin{eqnarray}
\frac{a^{\prime \prime}}{a} - \frac{a^{\prime^2}}{a^2} + \frac{8
\pi G}{3} \left( \phi^{\prime^2} - a^2 V(\phi) \right) &=& 0 \,,
\label{phia1}\\[2mm]
\phi^{\prime \prime} + 2 \frac{a^\prime}{a} \phi^{\prime} + a^2
\frac{\partial V(\phi)}{\partial \phi} &=& 0 \,.
\label{phia2}
\end{eqnarray}

Most of the contribution in the perturbation spectrum comes from
long-wavelength modes, since at late times they are non relativistic
and act like a classical homogeneous field with an amplitude $\bar\mu$
given by:
\begin{equation}
\bar{\mu}= \sqrt{ \langle \mu^2 \rangle} = \left(\frac{1}{2 \pi a^3} \int dn
\,n^2\,|\mu_n|^2 \right)^{1/2}\,.
\label{mubar}
\end{equation}
These are produced at early stages of inflation, thus they are very
sensitive to the initial conditions. The correct vacuum
state\footnote{In \cite{brand} the authors argue that there are two
vacuum states. The argument extends to the criteria for choosing the
right vacuum out of the two. Here we show there is only one true vacuum
state which reduces to the Minkowski vacuum only at a certain limit.}
is given by the solution to Eq. (\ref{modeneff}) as clarified below.
But if the stage of 
inflation is very long, and the Hubble parameter $H$ is not changing
considerably, then a Minkowski-like vacuum is a good first order
approximation for the initial vacuum state ($\eta \rightarrow
-\infty$), with:
\begin{equation}
\mu_n(\eta) \simeq \frac{1}{\sqrt{2 n}} e^{-i n \eta} \,.
\end{equation}
However, if $H$ was greater at the early stages of inflation, before
the last 60 e-foldings, and/or the inflationary stage is short, then
one must solve the wave equation and find the solution that minimizes
the energy \cite{initial,birrell,linde}. This is the correct vacuum state of
the system. Otherwise, if one a priori chooses the Minkowski vacuum to
be the initial vacuum state describing the system, the resulting value
of $\langle \mu^2 \rangle$ is considerably underestimated as shown
rigorously by Felder et al. \cite{felder}.

As already shown in Ref. \cite{brand}, Eq. (\ref{modeneff}) represents
particle production in a time-dependent background. We will follow the
method of Bogoliubov transformation to determine the spectrum.  The correct
initial condition for the vacuum state is the solution to the equation
that minimizes the energy. Hence, if the 'time-dependent' background
goes asymptotically flat at late times, then in that limit the
wavefunction should behave as a plane wave:
\begin{equation}
\mu_n \rightarrow_{\eta \rightarrow - \infty}
\frac{1}{\sqrt{\Omega_n^{in}}} e^{-i \Omega^{in}_n \eta} \,.
\end{equation}
As it is well known on this scenario, in general at late times one has
a squeezed state due to the curved background that mixes positive and
negative frequencies. The evolution of the mode function $\mu_n$ at
late times fixes the Bogoliubov coefficients $\alpha_n$ and $\beta_n$,
\begin{equation}
\mu_n \rightarrow_{\eta \rightarrow +\infty} \frac{\alpha_n}{\sqrt{2
\Omega^{out}_n}} e^{-i\Omega^{out}_n \eta} + \frac{\beta_n}{\sqrt{2
\Omega^{out}_n}} e^{+i \Omega^{out}_n \eta}\,.
\end{equation}
with the normalization condition:
\begin{equation}
|\alpha_n|^2 - |\beta_n|^2 =1\,.
\end{equation}
In the above expressions, $\Omega^{in}_n$ and $\Omega^{out}_n$ denote
the asymptotic values of $\Omega_n$ when $\eta \rightarrow \mp
\infty$. 
The spectrum of particles per mode is then calculated with the
conventional Bogoliubov method \cite{birrell}. 
The number of particles created $n$ and their energy density $\rho$ are
calculated by the following expressions
\begin{eqnarray}
\langle n \rangle &=& \frac{1}{2 \pi^2 a^3} \int dn~ n^2 |\beta_n|^2
\,,\\ 
\langle \rho \rangle &=& \frac{1}{2 \pi^2 a^4} \int n dn \int n_{eff}
dn_{eff}~ |\beta_n|^2  \nonumber \\
&=& \frac{1}{2 \pi^2 } \int k dk \int \omega(k) d\omega~   
|\beta_k|^2 \,. \label{energy}
\end{eqnarray}
If the resulting Bogoliubov coefficient $\beta_k$ of the particles
produced has a (nearly) thermal distribution, we can conclude that the
CMBR in our problem is that of a (nearly) black body spectrum. We
introduce the class of Epstein functions as the family of dispersion
relations in Section 3, and derive 
the CMBR spectrum from the exact solutions to the evolution equations.

Meanwhile, for the special features of our choice of dispersion
relation, the modes at very high momenta but of ultra low frequencies 
$\omega(k)$ are frozen for as long as the Hubble expansion rate $H$ of
the Universe dominates over their frequency. We refer to that as the 
$tail$ of the dispersion graph. In Fig. (\ref{fig2}), for the
dispersed $\omega^2(k)$ vs. $k$, the tail corresponds to all the modes
beyond the point $k_H$, where $k_H$ is defined by the condition  
 $\omega^2(k_H)=H^2_0$, where $H_0$ is the Hubble rate
today. It then follows that the tail modes are still frozen at
present. We calculate the total energy of the particles by using
Eq. (\ref{energy}), as well as the frozen energy of the $tail$.
Thus the energy of the  tail is a contribution to 
the dark energy of the universe: up to present it has
the equation of state of a cosmological constant term.  However,
through the Friedmann equation, $H$ is a decreasing function of time
because until now it has been dominated by the energy density of
matter and radiation. Therefore, whenever $H$ drops below the
frequency $\omega$ of 
an ultralow frequency mode, this mode becomes dynamic by picking a
kinetic term and redshifts away very quickly.  Hence, when the
dominant contribution to the evolution equation for $H$ comes from the 
$tail$ energy, the behaviour of those modes with equations of motion
coupled to the Friedmann equation becomes very complex. It is hard to
calculate at which rate $H$ drops in this situation. If
eventually, $H$ drops all the way to zero, all the modes in the tail
must have decayed. Their equation of state, when $H$ becomes zero,
is that of radiation. The reason can be traced back at their origin in
transplanckian regime. It is well known that scalar perturbations
produced during inflation do not contribute to the total energy. Thus
the origin of this modes is in the tensor perturbations.  In their
physical nature they correspond to gravitational waves of very short
distance but ultralow energy\footnote{We thank S. Carroll for pointing this
out.}. We calculate their energy today in Section 4.

We would like to elaborate on yet another possibility, which has not
been mentioned before in the literature, that can give rise to a
similar dispersion relation: very short or very large distance physics
may have a curvature different from the FLRW element of
Eq. (\ref{frw}), e.g. a different scale factor.  This becomes
clearer when recognizing the strong relation between the
time-dependent dispersion relation and the curvature given by the time
derivatives of the scale factor $a(\eta)$. The basic argument is the
observation that the modulated frequency  $\Omega_n^2$ in the wave
equation contains  the contribution of these two terms, as 
given in Eq. (\ref{omegan}). Therefore, while keeping the generalised
frequency invariant, changing the first term in 
$\Omega_n^2$  can be viewed or attributed to changes in the
second term, such that:
\begin{equation}
\Omega_n^2=a^2\omega_{nonlinear,k}^2 -(1-6 \xi)\,\frac{a^{\prime \prime}}{a} =
{\mathcal{A}}^2 \omega_{linear,k}^2 - (1-6 \xi)\,\frac{\mathcal{A}^
{\prime \prime}}{\mathcal{A}}\,,
\end{equation}
where $\mathcal{A}$ is the new scale factor at very short distances.
Even in the conformal case $\xi=1/6$, when the term $a^{\prime
\prime}/a$ drops out of Eq. (\ref{modeneff}), the time dependent
frequency $\omega_{nonlinear,k}^{2}$ can mimic a term proportional to
$a^{\prime \prime}/a$ at $\lambda(\eta)k_C \ll 1$.
Thus, any modulation of the dispersion relation is
equivalent to a change in the behavior of the time-dependence of the
background (a.k.a., the scale factor/curvature). In other words, we
could have introduced a different curvature at very short (or large)
distances from the start instead of a dispersed frequency\footnote{ 
We are using this equivalence in a sequencial paper \cite{paper} to
demonstrate that a different large scale curvature of the universe is
not possible as it conflicts with the observed CMBR data. Therefore,
trying to reinterpret the SN1a data in the light of a possible
different curvature for the large scale regions of the universe may be
ruled out.}.

\section{Exact Solution and the CMBR Spectrum}

We will consider the class of inflationary scenarios that through
Eqs. (\ref{phia1}) and (\ref{phia2}) has a power law solution for the scale
factor $a(\eta)$ in conformal time, 
$a(\eta)=|\eta_c/\eta|^\beta$, with $\beta \ge 1$, and the following
Epstein function \cite{epstein} for the dispersion relation:
\begin{eqnarray}
\omega^2(k) &=& F^2(k)= k^2 \left(\frac{\epsilon_1}{1+ e^x} +
\frac{\epsilon_2 e^x}{1+e^x} + \frac{\epsilon_3 e^x}{(1+e^x)^2}\right)
\,,\label{omega}\\ 
n^2_{eff} &=& a^2(\eta) F^2(n,\eta)= n^2 \left(\frac{\epsilon_1}{1+ e^x} +
\frac{\epsilon_2 e^x}{1+e^x} + \frac{\epsilon_3 e^x}{(1+e^x)^2}\right) \,,
\label{disp}
\end{eqnarray}
where $x=(k/k_C)^{1/\beta}=  A |\eta|$, with $
A=(1/|\eta_c|)(n/k_C)^{1/\beta}$. This is the most general expression
for this family of functions.  For our purposes, we will constrain
some of the parameters of the Epstein family in order to satisfy
the features required for the dispersion relation  as
follows.  
First, imposing the requirement of superstring duality, in order to have
ultralow frequencies for very high momenta, we
demand that the dispersion functions go asymptotically to zero. 
That fixes
\begin{equation}
\epsilon_2=0 \,.
\end{equation}
\begin{figure}[t]
\epsfxsize=10cm
\epsfxsize=10cm
\hfil \epsfbox{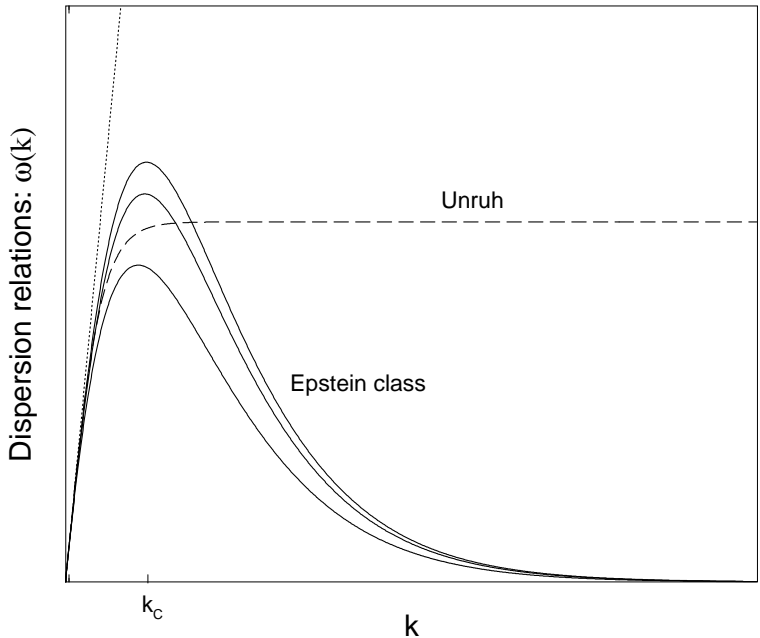} \hfil
\caption{{\footnotesize Shown is our family of dispersion relations,
for $\beta=1$ and representatives values of $\epsilon_1$ (solid
lines). We have also included the Unruh's 
dispersion relation (dashed line) and the linear one (dotted line) for
comparison.} }
\label{fig1}
\end{figure}
On the other hand, the condition of a nearly linear dispersion relation for
$k<k_C$ requires that
\begin{equation}
\frac{\epsilon_1}{2} + \frac{\epsilon_3}{4}=1 \,.
\end{equation}
Still we will have a whole family of functions parametrised by the
constant $\epsilon_1$, as can be seen in Fig. \ref{fig1}.

With the change of variables $\eta \rightarrow u= exp(A |\eta|)$, the
scalar wave equation (\ref{modeneff}) for the mode $\mu_n$ becomes:
\begin{equation}
\left[ \partial^2_u + \frac{1}{u} \partial_u+ V(u) \right] \mu_n =0
\,,
\label{modev}
\end{equation}
with:
\begin{equation}
V(u)=\frac{\hat\epsilon_1}{u^2(1+u)} + \frac{\hat\epsilon_3}{u
(u+1)^2} - \frac{( 1 - 6 \xi)}{u^2 A^2} \frac{a^{\prime \prime}}{a}
\label{potx}
\,,
\end{equation}
where:
\begin{equation}
\hat{\epsilon_i}= (k_C |\eta_c|)^2 \left(\frac{n}{k_C}\right)^{2(1 -
1/\beta)} \epsilon_i \,.
\end{equation}

In the case of conformal coupling to gravity, $\xi=1/6$,
Eq. (\ref{modev}) is exactly solvable in terms of hypergeometric
functions \cite{epstein}. This is a well studied case in the context
of particle creation in a curved background \cite{partcreation}. Even
if we are not in the case of conformal coupling, the contribution
\begin{equation}
\frac{a^{\prime \prime}}{a} = \frac{\beta (\beta+1)} {\eta^2} \,,
\end{equation}
is going to be negligible at early times ($\eta \rightarrow -\infty$); 
at late times, it can be absorbed in the dispersion relation
Eq. (\ref{disp}) redefining the constants $\epsilon_i$.  

As explained in Section 2, the correct initial condition is the
vacuum state solution that minimizes the energy. 
In the case where $\epsilon_2 \neq 0$, this vacuum state behaves as a
plane wave in the asymptotic limit $\eta \rightarrow -\infty$, with
$\Omega^{(in)}_n \rightarrow \sqrt{\epsilon_2} n$. 
However, when $\epsilon_2=0$ as in our case, 
the correct behavior of the mode function in the
remote past is given by the solution of Eq. (\ref{modeneff}) in the
limit $\eta \rightarrow -\infty$. The exact solution which matches
this asymptotic behavior is then given by:
\begin{equation}
\mu^{(in)}(\eta)= C^{in} \left(\frac{1
+u}{u}\right)^{d}~ _2F_1 [\frac{1}{2}+ d + b,\frac{1}{2}+ d-b, 1+2 d,
\frac{1 +u}{u}] \,, 
\end{equation} 
where $C^{in}$ is a normalization constant, and  
\begin{eqnarray}
b &=& i \tilde{b} = i \sqrt{ \hat{\epsilon}_1}  \,,\\ 
d &=& i \tilde{d} = \sqrt{\frac{1}{4} + \hat\epsilon_3} \,. 
\end{eqnarray}
At late times the
solution becomes a squeezed state by mixing of positive and negative
frequencies:
\begin{equation}
\mu_n \rightarrow_{\eta \rightarrow +\infty} \frac{\alpha_n}{\sqrt{2
\Omega^{out}_n}} e^{-i\Omega^{out}_n \eta} + \frac{\beta_n}{\sqrt{2
\Omega^{out}_n}} e^{+i \Omega^{out}_n \eta} \,,
\end{equation}
with $|\beta_n|^2$ being the Bogoliubov coefficient equal to the
particle creation number per mode $n$, and $\Omega^{out}_n \simeq
\sqrt{\epsilon_1} n$. Using the linear transformation
properties of hypergeometric functions \cite{abramowitz}, 
we find that\footnote{In the most general case, where $\epsilon_2 \neq 0$
in Eq. (\ref{disp}), it is obtained \cite{epstein}:
\begin{equation}
\left| \frac{\beta_n}{\alpha_n} \right| = 
\left| \frac{ \cosh \pi(\tilde{d} + \tilde{b}-\tilde{a})}{ \cosh
\pi(\tilde{d} - \tilde{b}-\tilde{a})} \right| \,, \label{bogol}
\end{equation}
with $\tilde{a}=\sqrt{\hat{\epsilon_2}}$. Also in this case the
spectrum of the fluctuations is nearly thermal, with the parameter
$\tilde{d}$ controlling the deviation
from thermality.}
\begin{eqnarray}
\left| \frac{\beta_n}{\alpha_n} \right| &=& e^{-2 \pi 
\tilde{b}} \left|\frac{ \cosh \pi(\tilde{d} + \tilde{b})}{ \cosh
\pi(\tilde{d} - \tilde{b})} \right| \label{betak}\,.
\end{eqnarray}
If $d$ is a real number ($\epsilon_3^\prime > -1/4$), then we
obtain:
\begin{equation}
|\beta_n|^2 = \frac{e^{-2\pi \tilde{b}}}{2 \sinh 2\pi \tilde{b}}
\label{betak2}\,.
\end{equation}
It is clear from Eqs. (\ref{betak}) and (\ref{betak2}) that the 
spectrum of created particles is nearly thermal to high 
accuracy\footnote{ We remind the reader that we have neglected the  
backreaction effects during the calculation, based on the result of a
small particle number per mode, in the high momentum regime ($k \gg M_P$) 
and a very small energy contained in these
modes. Clearly, the particle number per mode being small is consistent
with the result of the exponentially  suppressed, near-thermal Bogoliubov
coefficient. See Section 4 for the energy.},
\begin{equation}
|\beta_n|^2 \simeq e^{-4 \pi \tilde{b}} \,.
\end{equation} 
Thus, we can immediately conclude that the CMBR spectrum is that of a
(nearly) black body spectrum. That means the spectrum is (nearly)
scale invariant, i.e., the spectral index is $n_s \simeq 1$.
This is consistent with previous results obtained in the
literature \cite{brand, niemeyer, kowalski, kempf}, when using
a smooth dispersion relation and the correct choice of the initial
vacuum state, as discussed above. In Refs. \cite{brand} and
\cite{niemeyer}, dispersion relations, that were originally
applied to black hole physics \cite{jacobson,unruh}, were
used in the context of cosmology. New models of dispersion relations
were proposed by the authors of Refs. \cite{kowalski} and \cite{kempf}.
Our proposal for a 1-parameter class of models has a significantly
different feature from the above, namely:
the appearence of ultra-low frequency
modes in the transplanckian regime. The implications of such a behavior
for high momenta on the production of dark energy are discussed in the
next section. 

\section{Dark Energy from the ``Tail''}

In Section 2 we defined the $tail$ as the range of those modes in
the frequency dispersion class (originating from 
the transplanckian regime), whose frequency is less or at  most
equal to the present Hubble rate, $H_0$ (see Fig. 2). It then follows
that they have not 
decayed and redshifted away but are still frozen $today$.  Since $H$
has been a decreasing function of time, many modes, even those in the
ultralow frequency range, have become dynamic and redshifted away one
by one, everytime the above condition is broken, i.e. when the
expansion rate $H$ dropped below their frequency. Clearly, the other
modes have long decayed into radiation and the tail modes are the
only modes still frozen. They
contain vacuum energy of very short distance, hence of very low
energy. The last mode in the tail would decay when and if $H=0$. When
the tail modes become dynamic by acquiring a kinetic term (when
$\omega(k)>H$), they decay away as gravitational waves (explained in
Section 2).  The  $tail$ starts from some value $k_H$ which must be found by
solving the equation
\begin{equation}
\omega^2(k_H)=H^2_0 \,.
\label{omegah}
\end{equation}
The range of the modes defining the  $tail$ is then for $k_H<k<\infty$.
Their time-dependent behaviour when they decay depends on the
evolution of $H$ and is complicated because they contribute to the
expansion rate for $H$. Thus, their equations of motion are coupled to
the Friedmann equation.

\begin{figure}[t]
\epsfxsize=10cm
\epsfxsize=10cm
\hfil \epsfbox{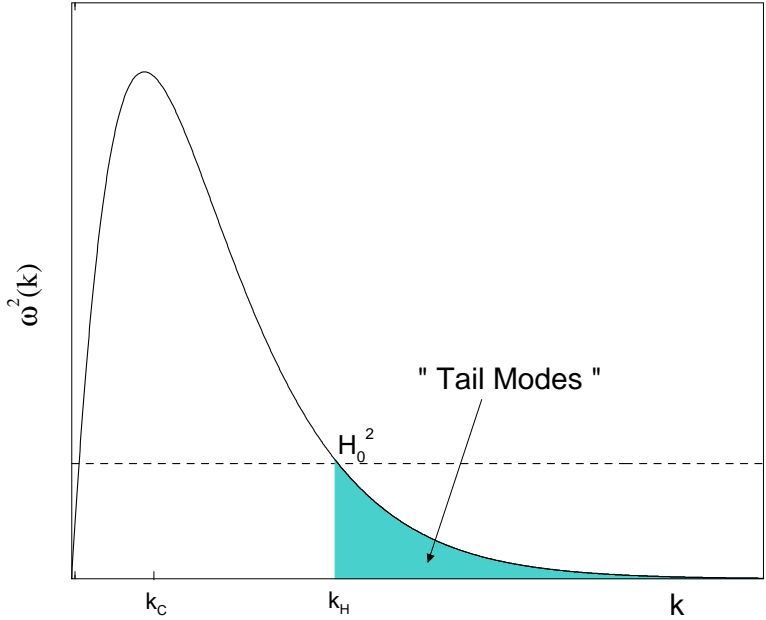} \hfil
\caption{{\footnotesize The range of modes in the tail, $k_H < k <
\infty$, defined by Eq. (\ref{omegah}). $H_0$ is the present value of
the Hubble constant.} }
\label{fig2}
\end{figure}

However we can calculate their contribution to the dark energy today,
when they are still frozen, thereby mimicking a cosmological
constant. We calculate numerically (using 'Mathematica') the range of
the modes in the tail from Eq. (\ref{omegah})  and use this value for
the limit of integration in the tail energy given by
Eq. (\ref{energy}). Below we report these results for the 
case of a scale factor with $\beta=1$ but 
other values of $\beta$ were also considered numerically and they
produce an even  smaller dark energy due to the extra suppression in
the integral coming from the Bogoliubov coefficient $\beta_k^{2}$.
Eq. (\ref{omegah}) is a messy transcendental equation but the solution
to that equation is crucial to the dark energy since the value $k_H$ is the
limit of the energy integral. That is why we solved Eq. (\ref{omegah})
numerically and replaced it in Eq. (\ref{energy}) for the energy,
using different representative values of the parameter $\epsilon_1$.

The energy for the tail is given by:
\begin{equation}
\langle \rho_{tail} \rangle = \frac{1}{2 \pi^2 } \int^{\infty}_{k_H} k dk
\int \omega(k) d\omega~ |\beta_k|^2  \label{entail}\,,
\end{equation}
while the expression for the total energy is:
\begin{equation}
\langle \rho_{total} \rangle = \frac{1}{2 \pi^2 } \int^{\infty}_{0} k dk 
\int \omega(k) d\omega~|\beta_k|^2 \,.
\end{equation}
The numerical calculation of the tail energy produced the following
result: for random  different values of 
the free parameters, the dark energy of the tail is
$\rho_{tail}=10^{-122} f(\epsilon_1)$, times less than the total energy
{\em during inflation},
i.e. $\frac{\rho_{tail}}{\rho_{total}}=10^{-122} f(\epsilon_1)$ at 
Planck time. The prefactor $f(\epsilon_1)$, which depends weakly on the
parameter of the dispersion family $\epsilon_1$, is a small number
between 1 to 9, which clearly can contribute at the most by 1 order of
magnitude. 

{\em This is an amazing result!} 
It can readily be checked by plugging in the dispersion expression
for $\omega_k$, Eq. (\ref{omega}), in the integral expression of
Eq. (\ref{entail}) for the tail energy, then
using as the limit 
of integration the value $k_H$ found by the condition in
Eq. (\ref{omegah}). 
This result  can be understood qualitatively by noticing that the 
behavior of the frequency for the ``tail'' modes is nearly an
exponential decay (see Eq. (\ref{omega})), and as such dominates over the
other terms in the energy integrand of Eq. (\ref{entail}):
\begin{equation}
\omega^2(k > k_C) \approx exp(-k/k_C) \,,
\end{equation}  
Hence, due to the decaying exponential, the main contribution to the
energy integral in Eq.
(\ref{entail}) comes from the highest value of this exponentially
decaying frequency, which is the value of the integrand at the tail
starting point, $k_H \sim O(M_P)$, i.e., 
\begin{equation}
\langle \frac{\rho_{tail}}{\rho_{total}} \rangle \approx
\frac{k_H^2}{M_P^4} \omega^2(k_H) \approx \frac{H_0^2}{M_P^2}\approx
10^{-122} \,.  
\end{equation} 
Due to the physical requirement that the tail modes must have always been
frozen, the tail starting frequency $\omega(k_H)$ is then proportional to
the current value of Hubble rate $H_0$ (Eq. (\ref{omegah})). 

We suspect this result is generic for any scenario that features 
{\it ultralow frequencies} which  exponentially decay to zero at
very high momenta for two reasons. First, all the modes with an
ultralow frequency $\omega<H_0$ will be frozen and thus produce dark
energy. Secondly, their contribution to the energy may be small because 
of the following. Due to this kind of dispersion in the high momentum
regime, the phase space available for the ultra-low frequency modes with
$\omega(k>k_C)$ gets drastically reduced when compared to the phase
space factor in the case of a non-dispersive transplanckian regime,
controlling in this way these modes contribution to the energy density. 
The result for the tail energy also means that the
tail energy dominates today's 
expansion of the universe. Thus, at present, we can not tell a priori
the evolution of these modes and the time when they may become
dynamic. Only the solution to the equation for the modes coupled
(strongly at present) 
to the Friedmann equation would answer the question as to whether $H$ will
continue to decrease whith time. If that were the case, then these
tail modes would also eventually become dynamic and decay. However, we
calculated the equation of state for the limiting case when $H=0$. In
this case, all the modes in the tail are dynamic. 
The calculation of the energy density of the tail in the dynamic case
from Eq. (\ref{energy}) 
confirms that the tail decays in the form of radiation, as expected
since their physical nature is that of gravity waves of very short distances
(but ultralow energy), originating from tensor perturbations during
inflation.

The opposite case is also a possible outcome to the coupled
equations. It's possible that the frozen modes of the tail will prevent
$H$ from dropping further below, in which case these modes will never
decay. We have not solved these coupled equation yet, therefore we are
just speculating on the two possible outcomes of that calculation. The
solution is left for future work.

At present, these modes, originating from the transplanckian regime,
are behaving as dark energy of the same magnitude as the current
total energy in the universe.  This idea is then a leap forward to this
longstanding and challenging problem of dark energy, for at least two 
reasons: first, inspired by superstring duality, it is very plausible
to speak of scenarios with ultralow frequencies and very high
momenta. The tail modes, that are frozen at present, provide a good
candidate for the dark energy as our calculations show. Secondly,
although smooth dispersion functions that model the transplanckian
regime do not affect the CMBR spectrum, this regime still leaves its
imprints in the contribution to the energy of the universe. This is a
rich and currently underexplored area to consider with respect to the
cosmological constant mystery.

\section{Conclusions}

In this work we investigated two phenomenological aspects of
transplanckian physics: the issue of dark energy production, and
the sensitivity of the observed CMBR spectrum to the transplanckian
regime.  For this purpose, a family of dispersion relations is
introduced that modulate the high frequencies of the inflationary
perturbation modes at large values of the momenta $k$ for the
transplanckian regime. The smooth dispersion relations are chosen such
that the  frequency graph attenuates to zero at very high $k$, thereby
producing ultralow frequencies corresponding to very short distances,
but it is nearly linear for low values of $k$ up to the cuttoff scale
$k_C=M_{P}$.

We present the exact solutions to the mode equations and calculate the
spectrum through the method of Bogoliubov coefficients. The resulting
CMBR spectrum is shown to be (nearly) that of a {\em black body}. This
calculation lends strong support to the conjecture that smooth
dispersion relations which ensure an adiabatic time-evolution of the
modes produce a nearly scale invariant spectrum. Further, we elaborate
on the issue of the {\em initial conditions} to which the spectrum is
highly sensitive and show that there is {\em no ambiguity} in the
correct choice of the initial vacuum state. The only initial vacuum is
the {\em adiabatic vacuum} obtained by the solution to the mode
equation. 
On the other hand, we showed that the assumption of neglecting the
possible back reaction effects of the tail modes on the inflaton field
is reasonable and is justified by the result of the tail energy
calculation of Section 4. Also the Bogoliubov
coefficient obtained, Eq. (\ref{bogol}), is exponentially suppressed, so
back-reaction
does not become significant. We would also like to stress that due to
the dispersion class of functions chosen, defined in the whole range of
momenta from zero to infinity, the total
energy contribution of the modes produced is {\it finite}, without the
need of applying any renormalization/subtraction scheme. In a sense, the
regularization-renormalization procedure is encoded in the class of 
dispersion we postulate\footnote{Because of these two results, we
do not have the problems mentioned in Refs. \cite{tanaka,starobinsky}
when discussing trans-planckian physics.}.

The most exciting result of this work is the generation of dark
energy in the observed amount for the present universe \cite{dark}. 
This has its origin at the transplanckian regime, due to the presence
of  the dispersed {\em tail modes} with ultralow frequencies
equal or less than the current Hubble constant. 
The evolution of these modes is given by their
equation of motion, and it depends on the Hubble rate through
the friction-like term $3 H_0 \dot \mu_n$. On the other hand, the
evolution of the Hubble rate, given by the Friedmann equation, contains
contribution from the energy of these modes. But currently, the
Hubble constant dominates over their frequency in the mode equation of
motion. It then follows that the tail modes, up to present, are still frozen
and have been behaving like a cosmological constant term.  Therefore
their energy is dark energy\footnote{For a different mechanism of 
generating a constant energy density, through the backreaction of
cosmological perturbations, that mimics a cosmological constant
term, see Refs. \cite{Tsamis,Brand}.}.

We have calculated numerically the energy of the $tail$ during the
inflationary stage for different values of the dispersion parameter
$\epsilon$. The calculation showed that the tail energy was
$10^{-122}$ (times a prefactor $f(\epsilon_1)$ which weakly depends
on  $\epsilon_1$ and, for random values of the parameter, takes values
between 1 to 9) orders less than the total energy during 
inflation. This result is true for the whole class of dispersion
relations. We chose different random values of the 1-parameter
dispersion family $\epsilon_1$, and the numerical calculation shows that
$\epsilon_1$ influences the energy at the most by less than an order of
magnitude. We {\em did not need to do any tuning}  of the parameters
and used the {\em  Planck scale as the fundamental scale of the
theory}. Clearly, at present the tail energy dominates in the Friedmann
equation, if the ratio of its energy to the total energy (as found by
the calculation) was $10^{-122} f(\epsilon_1)$ during inflation. The
$tail$ thus 
provides a {\em strong candidate}  for explaining the dark energy of
the universe \cite{dark}. We suspect that the above result of producing
such an extremely small number for the $tail$ energy without any fine-tuning
(and by using $M_{P}$ as the only fundamental scale of the theory), is
generic for any dispersion graph with a $tail$. The family of
dispersion relations that feature a $tail$, corresponding to vacuum
modes of very short distances, was motivated by superstring
duality \cite{superstring1,superstring2}. 

However, in Section 2 we made the observation that {\em
introducing a dispersion relation is equivalent to introducing changes
in the curvature of the 
universe}, at very short or very large distances while keeping the
generalized frequency $\Omega_n$ of Eq. (\ref{modeneff})
invariant. It is quite  
possible that the dispersion relation for the $tail$ modes results
from a different curvature of the universe at very short
distances. This is an important link and we use it in a sequential
paper \cite{paper} to demonstrate that the SN1a data can not be
reinterpreted away by changing the large scale curvature of the
universe. Although it is counterintuitive, since large distance would
correspond to low energy theories, we show in \cite{paper} that
any changes in large scale curvature would disagree with the observed
CMBR spectrum.

It would be interesting to know what happens to the tail and the
Hubble rate in the future. After all, a model is useful in so far as
it can make future predictions.
Although conceptually it is straightforward to find out the answer,
given by the solution to the coupled equations 
(\ref{modeneff}), (\ref{phia1}) and (\ref{phia2}), technically it
appears difficult to predict  
the future evolution of the tail modes. The technical difficulty lies 
in the fact that at present, the $tail$ equation of motion is 
strongly coupled to the Friedmann equation for $H$ since the $tail$ 
energy dominates. The Hubble rate  would continue to decrease {\em 
only if} these modes decay, but these modes can decay {\em only when} 
the Hubble rate decreases below their frequency. It may be 
possible that the frozen tail will sustain a constant Hubble rate 
which in turn will not allow the further decay of the tail modes. It 
is also possible that $H$ will continue to decrease in which case the 
tail modes will become dynamic and redshift away in the future. It is 
only the solution to the mode equations coupled to the Friedmann 
equation, that will provide the answer on whether the Hubble constant 
and tail will decay in the future or remain at their current 
value. We do not have this solution  yet, 
and the work is left for future investigation. In addition, the
equation of state, $w(t)= \langle p/\rho \rangle$, is an
$observable$ that will provide a test to the model \cite{turner},
especially with the new data coming in the near future from the SNAP
\cite{snap} and SDSS \cite{sdss} missions.  

However, we know that currently these $tail$ modes are frozen vacuum modes
of ultralow energy but very short distance, thus their energy behaves
like a cosmological constant energy. We also know they become dynamic
and acquire a kinetic term only when the Hubble rate drops below the  
frequency. And, if they decay,  the product is radiation of
gravity waves at very short distances since their physical origin is
from the tensor perturbations during inflation (it is well known that
scalar perturbations do not contribute to the energy). The condition
for the decay of the last mode in the tail is fulfilled when $H$ has
dropped to zero.

Many of our results, e.g. the dispersion family and the exact
solutions together with the Bogoliubov coefficients, could be applied to
the Black Hole Physics.  
The issue of transplanckian physics was 
originally raised in the Black Hole  context with respect to the
sensitivity of the
Hawking radiation to the blueshifted, superplanckian energy wavepackets.
Following a phenomenological approach,  
a few dispersive models were introduced \cite{jacobson}-\cite{jacobson2} in
order to introduce a bound on the blueshifted energies and check the
sensitivity of the black hole spectrum.   
We have introduced a new, different family of dispersive models, that
also gives rise to a thermal spectrum. The analytical results of our
class of dispersion models can be applied to the black
hole physics and reproduce the same thermal Hawking spectrum. 
There are many subtleties involved due to the different symmetries of the
two scenarios, but these issues are beyond the purpose and scope of this
paper. It is left for future work. However, if the Hawking radiation for
this class of dispersions is again thermal, it
lends  strong support to Unruh's conjecture that black hole radiation 
is insensitive to physics in the far ultraviolet
(trans-planckian) regime, being predominantly an infrared effect.
On the other hand, our class of models departs from the previous ones in
the 
asymptotic behavior at very high momenta, with the presence of an 
infinite ``tail'' of ultralow frequency modes.    
The $tail$ feature and energy results, applied to a black hole 
case, may raise interesting issues, in particular with respect to the
black hole's information loss paradox.

{\bf Acknowledgment:} We are very grateful to A. Riotto for many
beneficial comments and for pointing out the connection with superstring
duality. We also want to thank R. Barbieri, R. Rattazzi, L. Pilo,
S. Carroll, C. T. Hill for helpful discussions. P.K. would like to
acknowledge financial support by EC under the TMR contract
No. HPRN-CT-2000-00148.

\end{document}